\documentclass[12pt, amsmath, amssymb]{revtex4}

\usepackage{amsmath,amssymb,amsthm,amsfonts}
\usepackage{enumerate}
\usepackage{epsfig}
\usepackage{graphicx}
\usepackage{pgf}

\usepackage{natbib}
\bibpunct{}{}{,}{s}{,}{,}
\makeatletter
\DeclareRobustCommand\onlinecite{\@onlinecite}
\def\@onlinecite#1{\begingroup\let\@cite\NAT@citenum\citealp{#1}\endgroup}
\makeatother 

\newtheorem{theorem}{Theorem}[section]
\newtheorem*{theorem*}{Theorem}

\newtheorem*{corollary*}{Corollary}

\newtheorem*{conjecture*}{Conjecture}

\theoremstyle{remark}

\theoremstyle{definition}
\newtheorem{definition}{Definition}

\newtheorem{example}{Example}
\newtheorem{algorithm}{Algorithm}

\def\N{\mathbb N_{\ge 0}}

\begin{document}

\title{A modified Next Reaction Method for simulating chemical systems
  with time dependent propensities and delays}

\author{David F. Anderson}%
\email{anderson@math.wisc.edu}
\homepage{www.math.wisc.edu/~anderson}
\affiliation{Department of Mathematics, University of
  Wisconsin-Madison, Madison, Wi 53706}

\date{\today}

\begin{abstract}
  Chemical reaction systems with a low to moderate number of molecules
  are typically modeled as discrete jump Markov processes.  These
  systems are oftentimes simulated with methods that produce
  statistically exact sample paths such as the Gillespie Algorithm or
  the Next Reaction Method.  In this paper we make explicit use of the
  fact that the initiation times of the reactions can be represented
  as the firing times of independent, unit rate Poisson processes with
  internal times given by integrated propensity functions.  Using this
  representation we derive a modified Next Reaction Method and, in a
  way that achieves efficiency over existing approaches for exact
  simulation, extend it to systems with time dependent propensities as
  well as to systems with delays.
\end{abstract}

\maketitle

\section{Introduction}

Due to advances in the knowledge of cellular systems, where there are
low to moderate numbers of molecules of certain species, there has
been a renewed interest in modeling chemical systems as discrete and
stochastic as opposed to deterministic and continuous.
\cite{Arkin1998, McAdams1997, Ozbudak2002, Petzold2005} Because of the
intrinsic stochasticity at this level, understanding of a given system
is gained through knowledge of the distribution of the state of the
system at a given time.  As it is typically impracticable to
analytically solve for the distribution of the state of the system at
a particular time for all but the simplest of examples, simulation
methods have been developed that generate statistically exact sample
paths so as to approximate the distribution.  The two most widely used
exact simulation methods are the original Gillespie Algorithm
\cite{Gill76, Gill77} and the Next Reaction Method of Gibson and
Bruck. \cite{Gibson2000}

In this paper, we will explicitly represent the reaction times of
discrete stochastic chemical systems as the firing times of
independent, unit rate Poisson processes with internal times given by
integrated propensity functions.  Such a representation is not novel
and is called a random time change representation in the mathematics
literature.  See, for example, Refs. \onlinecite{Kurtz80, Kurtz86,
  Ball06}.  However, using such a representation in an explicit
attempt to develop new simulation methods has a number of benefits
that have seemingly not been explored in the chemistry literature.
First, the representation will naturally lead us to a modified version
of the Next Reaction Method. \cite{Gibson2000} Second, the modified
Next Reaction Method will be shown to be the natural choice for
simulating systems with propensities that depend explicitly on time
(such as systems with variable temperature or cell volume).  Third, we
will be able to easily extend our modified Next Reaction Method to
systems that allow delays between the initiation and completion of
reactions in a manner that achieves efficiency over existing methods.
More precisely, in our modified Next Reaction Method for systems with
delays no random numbers or computations will be wasted (such as
happens in the method of Bratsun et al. \cite{Bratsun2005} and Barrio
et al.  \cite{Barrio2006}) (see Section \ref{sec:delays}), and there
will be no need for the complicated machinery of the method developed
by Cai \cite{Cai2007} (see Section \ref{sec:delays}) in the handling
of the stored delayed reactions.  We note that the ideas we use to
develop our modified Next Reaction Method are analogous to the
theories of generalized semi-Markov processes \cite{Burman81,
  Schass78, Glynn89} and stochastic Petri nets \cite{Haas2002}, and
can also be extended to develop new accurate and efficient approximate
tau-leaping methods. \cite{Anderson2007b}

The outline of the paper is as follows. In Section \ref{sec:gill} we
briefly present the original Gillespie Algorithm.  In Section
\ref{sec:modeling} we introduce our representation of the reaction
times as the firing times of independent, unit rate Poisson processes
with internal time given by integrated propensity functions.  In
Section \ref{sec:nrm} we rederive the Next Reaction Method and derive
a modified Next Reaction Method using the representation detailed in
Section \ref{sec:modeling}. In Section \ref{sec:time-dep} we consider
systems with propensities that depend explicitly on time and conclude
that our modified Next Reaction Method is the preferable algorithm to
use in such cases.  In Section \ref{sec:delays} we consider systems in
which there is a delay between the initiation and completion of some
of the reactions and develop a new algorithm for simulating such
systems that is an extension of our modified Next Reaction Method.

\section{The Gillespie Algorithm}
\label{sec:gill}

Consider a system consisting of $N \ge 1$ chemical species,
$\{X_1,\dots,X_N\}$, undergoing $M \ge 1$ chemical reactions, each of
which is equipped with a propensity function (or intensity function in
the mathematics literature), $a_k(X)$. For the time being, assume that
the time between the initiation and the completion of each reaction is
negligible. To accurately simulate the time evolution of the number of
each species, $X(t) = \{X_1(t),\dots,X_N(t)\} \in \N^N$, one needs to
be able to calculate 1) how much time will pass before the next
reaction takes place (i.e. initiates and completes) and 2) which
reaction takes place at that future time.  One can then simulate
statistically exact sample paths for the system of interest.  The
following assumption, sometimes called the fundamental premise of
chemical kinetics, is based upon physical principles and serves as the
base assumption for simulation methods of chemically reacting systems:
\cite{Gill77}
\begin{align}
  \begin{split}
    a_k(X(t))\Delta t + o(\Delta t) =  \ &\hbox{the probability that
      reaction $k$}\\ 
    &\hbox{takes place in a small time interval $[t, t + \Delta t)$},
    \label{eq:FundPrem}
  \end{split}
\end{align}
where $o(\Delta t)/\Delta t \to 0$ as $\Delta t \to 0$.  Based upon
the assumption \eqref{eq:FundPrem}, the time until the next reaction,
$\Delta$, is exponentially distributed with parameter $a_0(X(t)) =
\sum_{k=0}^M a_k(X(t))$ and the probability that the next reaction is
the $k$th is $a_k(X(t))/a_0(X(t))$. These observations form the
foundation for the well known Gillespie Algorithm. \cite{Gill76,
  Gill77}

\begin{algorithm}\textbf{(Gillespie Algorithm)}
  \begin{enumerate}
    \setlength{\itemsep}{1pt}
    \setlength{\parskip}{0pt}
    \setlength{\parsep}{0pt}
  \item Initialize.  Set the initial number of molecules of each
    species and set $t = 0$.
  \item Calculate the propensity function, $a_k$, for each reaction.
  \item Set $a_0 = \sum_{k=1}^M a_k$.
  \item Generate two independent uniform(0,1) random numbers $r_1$ and
    $r_2$.
  \item Set $\Delta = 1/a_0\ln(1/r1)$ (equivalent to drawing an
    exponential random variable with parameter $a_0$).
  \item Find $\mu \in [1,\dots,M]$ such that
    \begin{equation*}
      \sum_{k=1}^{\mu-1} a_k < r_2a_0 \le \sum_{k=1}^{\mu} a_k,
    \end{equation*}
    which is equivalent to choosing from reactions $[1,\dots,M]$ with
    the $k$th reaction having probability $a_k/a_0$.
  \item Set $t = t + \Delta$ and update the number of each molecular
    species according to reaction $\mu$.
  \item Return to step 2 or quit.
  \end{enumerate}
  \label{alg:gill}
\end{algorithm}


We point out that the Gillespie Algorithm uses two random numbers per
step.  The first is used to find {\em when} the next reaction occurs
and the second is used to determine {\em which} reaction occurs at
that time.  In Section \ref{sec:nrm} we will demonstrate how the Next
Reaction Method generates exact sample paths while only needing one
random number per step.

\section{Representation using Poisson processes}
\label{sec:modeling}

We will explicitly represent the reaction times of chemical systems as
the firing times of Poisson processes with internal times given by
integrated propensity functions. \cite{Kurtz80, Kurtz86, Ball06} Using
such a representation allows us to consider the system as a whole via
a stochastic integral equation as opposed to solely considering how to
calculate when the next reaction occurs and which reaction occurs at
that time.  The benefits of such a representation will stem from the
fact that the randomness in the model is separated from the state of
the system.

Let $\nu_k, \nu_k' \in \N^n$ be the vectors representing the number of
each species consumed and created in the $k$th reaction, respectively.
Then, if $R_k(t)$ is the number of times that the $k$th reaction has
taken place up to time $t$, the state of the system at time $t$ is
\begin{equation}
  X(t) = X(0) + \sum_{k=1}^M R_k(t)(\nu_k' - \nu_k).
  \label{eq:xR}
\end{equation}
However, based upon the assumption \eqref{eq:FundPrem}, $R_k(t)$ is a
counting process with intensity $a_k(X(t))$ such that Prob$(R_k(t +
\Delta t) - R_k(t) = 1) = a_k(X(t))\Delta t$ for small $\Delta t$.
Therefore,
\begin{equation}
  R_k(t) = Y_k\left(\int_0^t a_k(X(s)) ds\right),
  \label{eq:Rk}
\end{equation}
where the $Y_k$ are independent, unit rate Poisson processes.  Thus,
$X(t)$ can be represented as the solution to the following equation:
\begin{equation}
  X(t) = X(0) + \sum_{k=1}^M Y_k\left(\int_0^t a_k(X(s))
    ds\right) (\nu_k' - \nu_k).
  \label{eq:Main}
\end{equation}
Note that the state of the system, $X(s)$, and hence each propensity
function $a_k(X(s))$, is constant between reaction times.  In Section
\ref{sec:time-dep} we will consider systems in which the propensity
functions are not constant between reactions, such as arise due to
changes in temperature or cellular volume.

We make two points that are crucial to an understanding of how
different simulation methods arise from equation \eqref{eq:Main}.
First, all of the randomness in the system is encapsulated in the
$Y_k$'s and has therefore been separated from the state of the system.
Thus, since the system \eqref{eq:Main} only changes when one of the
$Y_k$'s change, the relevant question of each simulation algorithm is
how to efficiently calculate the firing times of each $Y_k$ and how to
translate that information into reaction times for the chemical
system.  Second, there are actually $M+1$ relevant time frames in the
system.  The first time frame is the actual, or absolute time, $t$.
However, each Poisson process $Y_k$ brings its own time frame. More
specifically, if we define $T_k(t) = \int_0^t a_k(X(s)) ds$ for each
$k$, then it is relevant for us to consider $Y_k(T_k(t))$. We will
call $T_k(t)$ the ``internal time'' for reaction $k$.  
\begin{definition}
  For each $k \le M$, $T_k(t) = \int_0^t a_k(X(s)) ds$ is the {\em
    internal time} of the Poisson process $Y_k$ of equation
  \eqref{eq:Main}.
\end{definition}
\noindent We will use the internal times of the system in an analogous
manor to the use of ``clocks'' in the theory of generalized
semi-Markov processes. \cite{Burman81, Schass78, Glynn89}

We now formulate the Gillespie Algorithm (Algorithm \ref{alg:gill}) in
terms of equation \eqref{eq:Main}. At time $t$, we know the state of
the system, $X(t)$, the propensity functions, $a_k(X(t))$, and the
internal times, $T_k(t)$.  Calculating 1) how much time will pass
before the next reaction takes place and 2) which reaction takes place
at that future time is equivalent to calculating 1) how much time
passes before one of the Poisson processes, $Y_k$, fires and 2) which
$Y_k$ fires at that later time.  Combining the previous statement with
the fact that the intensities of the Poisson processes are $a_k(X(t))$
yields one step of the Gillespie algorithm.  Use of the loss of memory
property for Poisson processes (which negates knowledge of the
internal times $T_k(t)$) allows us to perform subsequent steps
independently of previous steps.

Note that in the Gillespie Algorithm the firing times of the
individual processes $Y_k$ were calculated by first finding the time
required until any of them fired, and then calculating which reaction
fired at that future time.  In the next section we show how the Next
Reaction Method and our modified Next Reaction Method first calculates
when {\em each} of the $Y_k$ fires next, and then finds the specific
reaction that fires by taking the minimum of such times.  By not
invoking the loss of memory property (and, hence, differentiating
themselves from the First Reaction Method \cite{Gill77}), the Next
Reaction Method and modified Next Reaction Method make use of the
internal times $T_k(t)$ to nearly cut in half the number of random
variables needed per simulation.

\section{A modified Next Reaction Method}
\label{sec:nrm}

We again consider the system \eqref{eq:Main}.  At time $t$ we know the
state of the system $X = X(t)$, the propensity functions $a_k =
a_k(X(t))$, and the internal times $T_k = T_k(t)$.  We also assume
that we know $\Delta t_k$, the amount of absolute time that must pass
in order for the $k$th reaction to fire assuming that $a_k$ stays
constant over the interval $[t,t+\Delta t_k)$.  Therefore, $\tau_k = t
+ \Delta t_k$ is the time of the next firing of the $k$th reaction if
no other reactions fire first.  Note that if $t = 0$ and this is the
first step in the simulation of the system (and so $T_k = 0$), finding
each $\Delta t_k$ is equivalent to taking a draw from an exponential
random variable with parameter $a_k$.  Because we know $\Delta t_k$,
the internal time at which reaction $k$ fires is given by $T_k +
a_k\Delta t_k$.  In order to simulate one step, we now note that the
next reaction occurs after a time period of $\Delta = \min_k\{\Delta
t_k\}$, and the reaction that fires is the one for which the minimum
is achieved, $\mu$ say.  Therefore, we may update the system according
to reaction $\mu$, update the absolute time by adding $\Delta$ and
update the internal times by adding $a_k \Delta$ to $T_k$, for each
$k$.

For the moment we denote $\overline t = t + \Delta$ and the updated
propensity functions by $\overline a_k$.  The relevant question now
is: for each $k$, what is the new absolute time of the firing of
$Y_k$, $\overline \tau_k$, assuming no other reaction fires first?
For reaction $\mu$, we must generate its next firing time from an
exponential random variable with parameter $\overline a_k$.  For $k
\ne \mu$ we note that, in general, the new absolute firing times will
not be the same as the old because the propensity functions have
changed.  However, the {\em internal time} of the next firing of $Y_k$
has not changed and is still given by $T_k(t) + a_k\Delta t_k$.  We
also know that the updated internal time of $Y_k$ is given by
$T_k(\overline t) = T_k(t) + \Delta a_k$.  Therefore, the amount of
internal time that must pass before the $k$th reaction fires is given
as the difference
\begin{equation*}
  (T_k(t) + a_k\Delta t_k) - (T_k(t) + \Delta a_k) = a_k(\Delta t_k - \Delta).
\end{equation*}
Thus, the amount of absolute time that must pass before the $k$th
reaction channel fires, $\Delta \overline t_k$, is given as the
solution to $\overline a_k \Delta \overline t_k = a_k(\Delta t_k -
\Delta)$, and so
\begin{equation*}
  \Delta \overline t_k = \frac{a_k}{\overline a_k}(\Delta t_k - \Delta).
\end{equation*}
Thus, we see that
\begin{equation*}
  \overline \tau_k = \frac{a_k}{\overline a_k}(\Delta t_k - \Delta) +
  \overline t 
  = \frac{a_k}{\overline a_k}((t + \Delta t_k) - (t + \Delta)) + \overline t
  = \frac{a_k}{\overline a_k}(\tau_k - \overline t) + \overline t.
\end{equation*}
We have therefore found the absolute times of the next firings of
reactions $k \ne \mu$ without having to generate any new random
numbers.  Repeated application of the above ideas yields the Next
Reaction Method. \cite{Gibson2000}

\begin{algorithm} \textbf{(The Next Reaction Method)}
  \begin{enumerate}
    \setlength{\itemsep}{1pt}
    \setlength{\parskip}{0pt}
    \setlength{\parsep}{0pt}
  \item Initialize.  Set the initial number of molecules of each
    species and set $t = 0$.
  \item Calculate the propensity function, $a_k$, for each reaction.
  \item Generate $M$ independent, uniform(0,1) random numbers $r_k$.
  \item Set $\tau_k = 1/a_k\ln(1/r_k)$.
  \item Set $t = \min_k\{\tau_k\}$ and let $\tau_{\mu}$ be the time
    where the minimum is realized.
    \label{impstep1}
  \item Update the number of each molecular species according to
    reaction $\mu$.
  \item Recalculate the propensity functions for each reaction and
    denote by $\overline a_k$.
  \item For each $k \ne \mu$, set $\tau_k = (a_k/\overline a_k)(\tau_k
    - t) + t.$
    \label{step:8}
  \item For reaction $\mu$, let $r$ be uniform(0,1) and set
    $\tau_{\mu} = 1/\overline a_{\mu}\ln(1/r) + t$.
  \item For each $k$, set $a_k = \overline a_k$.
  \item Return to step \ref{impstep1} or quit.
  \end{enumerate}
  \label{alg:nrm}
\end{algorithm}


Note that after the first timestep is taken in the Next Reaction
Method, all subsequent timesteps only demand one random number to be
generated.  This is compared with two random numbers needed for each
step of the original Gillespie Algorithm (Algorithm 1).  We also note
that the Next Reaction Method was originally developed with the notion
of a dependency graph and a priority queue in order to increase
computational efficiency (see Ref. \onlinecite{Gibson2000} for full
details).  The dependency graph is used in order to only update the
propensities that actually change during an iteration (and thereby cut
down on unnecessary calculations) and the priority queue was used to
quickly determine the minimum value in Step \ref{impstep1}.  We have
omitted the details of these items as they are not necessary for an
understanding of the algorithm itself.  However, we point out that the
use of a dependency graph in order to efficiently update the
propensity functions is useful in any of the algorithms presented in
this paper, and not just to the Next Reaction Method.

We now present an algorithm that is completely equivalent to Algorithm
\ref{alg:nrm}, but makes more explicit use of the internal times $T_k$.
In the following algorithm, we will denote by $P_k$ the first firing
time of $Y_k$, in the time frame of $Y_k$, that is strictly larger
than $T_k$.  That is, $P_k = \min\{s > T_k : Y_k(s) > Y(T_k)\}$.  The
main idea of the following algorithm is that by equation
\eqref{eq:Main} the value
\begin{equation}
  \Delta t_k = (1/a_k)(P_k - T_k)
  \label{eq:tosolve}
\end{equation}
gives the amount of absolute time needed until the Poisson process
$Y_k$ fires assuming that $a_k$ remains constant.  Of course, $a_k$
does remain constant until the next reaction takes place.  Therefore,
a minimum of the different $\Delta t_k$ gives the time until the next
reaction takes place.  Thus, if we keep track of $P_k$ and $T_k$
explicitly, we can simulate the systems without the time conversions
of step \ref{step:8} of Algorithm \ref{alg:nrm}.

\begin{algorithm}\textbf{(Modified Next Reaction Method)}
  \begin{enumerate}
    \setlength{\itemsep}{1pt}
    \setlength{\parskip}{0pt}
    \setlength{\parsep}{0pt}
  \item Initialize.  Set the initial number of molecules of each
    species. Set $t = 0$.  For each $k$, set $P_k = 0$ and $T_k = 0$.
  \item Calculate the propensity function, $a_k$ for each reaction.
  \item Generate $M$ independent, uniform(0,1) random numbers $r_k$.
  \item Set $P_k = \ln(1/r_k)$.
    \label{modstep1}
  \item Set $\Delta t_k = (P_k - T_k)/a_k$.
    \label{impstep2}
  \item Set $\Delta = \min_k\{\Delta t_k\}$ and let $\Delta t_{\mu}$
    be the time where the minimum is realized.
  \item Set $t = t + \Delta$ and update the number of each molecular
    species according to reaction $\mu$.
  \item For each $k$, set $T_k = T_k + a_k\Delta$.
  \item For reaction $\mu$, let $r$ be uniform(0,1) and set $P_{\mu} =
    P_{\mu} + \ln(1/r)$.
    \label{modstep2}
  \item Recalculate the propensity functions, $a_k$.
  \item Return to step \ref{impstep2} or quit.
  \end{enumerate}
  \label{alg:modnrm}
\end{algorithm}


We note that Algorithms \ref{alg:nrm} and \ref{alg:modnrm} have the
same simulation speeds on all systems that we have tested.  This was
expected as the two are equivalent.  However, as will be shown in the
next section, Algorithm \ref{alg:modnrm} extends itself to systems
with time dependent rate constants in a smooth way, whereas Algorithm
\ref{alg:nrm} does not.  We also point out another nice quality of
Algorithms \ref{alg:nrm} and \ref{alg:modnrm}.  Suppose that a system
is governed by equation \eqref{eq:Main} except that the $Y_k$'s are no
longer Poisson processes.  That is, we suppose that the reactions do
not have exponential waiting times, but have waiting times given by
some other distribution.  To modify Algorithms \ref{alg:nrm} and
\ref{alg:modnrm} to handle such a situation, one solely needs to
change steps \ref{modstep1} and \ref{modstep2} in each so that the
waiting times are drawn from the correct distribution.

\section{Time dependent propensity functions}
\label{sec:time-dep}

Due to changes in temperature and/or volume, the rate constants of a
(bio)chemical system may change in time.  Therefore, the propensity
functions will no longer be constant between reactions.  That is,
$a_k(t) = a_k(X(t),t),$ and the full system is given by
\begin{equation}
  X(t) = X(0) + \sum_{k=1}^M Y_k\left(\int_0^t a_k(X(s),s)
    ds\right) (\nu_k' - \nu_k),
  \label{eq:timedep}
\end{equation}
where the $Y_k$ are independent, unit rate Poisson processes.  We
consider how to simulate system \eqref{eq:timedep} using the Gillespie
Algorithm, the Next Reaction Method, and our modified Next Reaction
Method.

\vspace{.125in}

\noindent \textbf{The Gillespie Algorithm.}  At time $t$ we know the
state of the system, $X(t)$, and, until the next reaction takes place,
the propensity functions $a_k(X(t),s)$, for $s > t$.  When the
propensity functions depended only on the state of the system the
Gillespie Algorithm calculated the time until the next reaction by
considering the first firing time of $M$ time-homogeneous Poisson
processes.  However, we now need to calculate the first firing time of
$M$ time-inhomogeneous Poisson processes.  It is a simple exercise to
show that the amount of time that must pass until the next reaction
takes place, $\Delta$, has distribution function
\begin{equation}
  1 - \exp \left( - \sum_{k = 1}^M \int_t^{t + \Delta} a_k(X(t),s) ds \right).
  \label{eq:gillvarying}
\end{equation}
Note that $X(t)$ is constant in the above integrals because no
reactions take place within the time interval $[t,t+\Delta)$.  Using
equation \eqref{eq:gillvarying}, $\Delta$ is found by first letting
$r$ be uniform$(0,1)$ and then solving the following equation:
\begin{equation}
  \sum_{k = 1}^M \int_t^{t + \Delta} a_k(X(t),s) ds = \ln(1/r).
  \label{eq:gillsolve}
\end{equation}
In Appendix \ref{app:A} we show that the reaction that fires at that
time will be chosen according to the probabilities
$a_k(X(t),t+\Delta)/a_0$, where $a_0 = \sum_{k=1}^M
a_k(X(t),t+\Delta)$.  Solving equation \eqref{eq:gillsolve} either
analytically or numerically will be extremely difficult and time
consuming in all but the simplest of cases.

\vspace{.125in}

\noindent \textbf{The Next Reaction Method.}  We begin by considering
the first step of the Next Reaction Method.  At time $t=0$, we need to
know the first firing times of independent, inhomogeneous Poisson
processes.  Therefore, we calculate the time that the $k$th reaction
channel will fire (assuming no other reaction fires first) by solving
for $\tau_k$ from:
\begin{equation}
  \int_0^{\tau_k} a_k(X(0),s) ds = \ln(1/r_k),
  \label{eq:nrmsolve}
\end{equation}
where $r_k$ is uniform$(0,1)$.  Equation \eqref{eq:nrmsolve} can be
solved either analytically or numerically.  Say that reaction $\mu$ is
the first to fire and does so at time $t$.  It is clear that to
calculate the next firing time of reaction $\mu$ we will need to
generate another uniform$(0,1)$ random variable $r_{\mu}$ and solve
\begin{equation*}
  \int_{t}^{\tau_{\mu}} a_k(X(t),s) ds = \ln(1/r_{\mu}).
\end{equation*}
What is less clear is how to reuse the information contained in
$\tau_{k}$ for $k \ne \mu$.

Proceeding as in Ref. \onlinecite{Gibson2000}, denote by $F_{n,a}$ the
distribution function for the $n$th firing of a reaction, where $a$ is
some parameter of the function.  Gibson and Bruck prove the following:

\begin{theorem}[Gibson and Bruck's generation of next firing time
  \cite{Gibson2000}]
  Let $\tau$ be a random number generated according to an arbitrary
  distribution with parameter $a_n$ and distribution function
  $F_{a_n,n}$.  Suppose the current simulation time is $t_n$, and the
  new parameter (after a step in the system in which this reaction did
  not fire) is $a_{n+1}$. Then the transformation
\begin{equation}
  \tau^* = F^{-1}_{a_{n+1},n+1}\left([F_{a_n,n}(\tau) -
    F_{a_n,n}(t_n)] / [1 - F_{a_n,n}(t_n)] \right) 
  \label{eq:nrmsolveTheorem}
\end{equation}
generates a random variable from the correct (new) distribution.  That
is, $\tau^*$ has the correct distribution of the next firing time.
\label{th:GB}
\end{theorem}

Gibson and Bruck demonstrate use of the above theorem on a system
whose volume is increasing linearly in time.  In this specific case,
it is possible to find closed form solutions of the distribution
functions and their inverses.  However, in general, calculating the
distribution functions and their inverses may be a difficult (or
impossible) task.  For this reason Gibson and Bruck conclude ``In
general, it (the above method) may not be at all practicable and it
may be easier to generate fresh random variables (according to the new
distribution function).''\cite{Gibson2000}

In a situation in which Theorem \ref{th:GB} is not practicable, the
following steps must be taken to move one timestep beyond time $t$.
First, generate uniform$(0,1)$ random variables, $r_k$, and then solve
\begin{equation}
  \int_{t}^{\tau_k} a_k(X(t),s) ds = \ln(1/r_k)
  \label{eq:nrmbad}
\end{equation}
for the time of the next firing of reaction $k$. We have been forced
to use the loss of memory property of Poisson processes and generate
new random variables.  Thus, we are really now performing the First
Reaction Method. \cite{Gill77}

\vspace{.125in}

\noindent \textbf{The modified Next Reaction Method.}  As above, we
begin by considering the first step of our modified Next Reaction
Method.  At time $t=0$, we set $T_k = 0$ and $P_k = \ln(1/r_k)$, where
each $r_k$ is uniform$(0,1)$.  To find the amount of time that must
pass before the $k$th reaction channel will fire if no others do first
we solve for $\Delta t_k$ from:
\begin{equation}
  \int_0^{\Delta t_k} a_k(X(0),s) ds = P_k - T_k = P_k.
  \label{eq:mnrmsolve}
\end{equation}
Again supposing that reaction $\mu$ fires first at time $t$, we update
$T_k = \int_0^{t} a_k(X(0),s) ds$ for each $k$.  In order to calculate
$\Delta t_{\mu}$ we must generate a new uniform$(0,1)$ random number,
$r_{\mu}$, set $P_{\mu} = P_{\mu} + \ln(1/r_{\mu})$ and solve
\begin{equation*}
  \int_t^{t + \Delta t_{\mu}} a_{\mu}(X(t),s) ds = P_{\mu} - T_{\mu}.
\end{equation*}
For $k \ne \mu$ we still know that $P_k$ is the internal time of the
next firing of reaction $k$, and so the amount of absolute time that
must pass, $\Delta t_k$, before the $k$th firing is given as the
solution to
\begin{equation}
  \int_{t}^{t + \Delta t_k} a_k(X(t),s)ds = P_k - T_k.
  \label{eq:solvemnrm}
\end{equation}

Therefore, by keeping track of the internal times $P_k$ and $T_k$ we
have been able to easily calculate the next firing of each reaction
without having to generate another random number.  We point out that
using equation \eqref{eq:solvemnrm} to solve for the next firing time
is no more difficult than using equation \eqref{eq:nrmbad} to find the
next firing time in the Next Reaction Method, but equation
\eqref{eq:nrmbad} demanded the generation of a random variable.  We
also point out that if there are closed form solutions to the above
integral, such as the case of linearly increasing volume, then this
method becomes very efficient.  Further, even in this case of time
dependent propensity functions, the modified Next Reaction Method
easily lends itself to situations in which the waiting times between
reactions are not exponential (only the generation of the $P_k$'s
changes).  We conclude that our modified Next Reaction Method will be
preferable to either the Gillespie Algorithm or the Next Reaction
Method on systems with propensity functions that depend explicitly on
time.

\section{Systems with delays}
\label{sec:delays}

We now turn our attention to systems in which there are delays,
$\tau_k > 0$, between the initiation and completion of some, or all,
of the reactions.  We note that the definition of $\tau_k$ has
therefore changed and is no longer the next reaction time of the Next
Reaction Method.  We partition the reactions into three sets, those
with no delays, denoted $ND$, those that change the state of the
system only upon completion, denoted $CD$, and those that change the
state of the system at both initiation and completion, denoted $ICD$.
The assumption \eqref{eq:FundPrem} becomes the following for systems
with delays:
\begin{align}
  \begin{split}
    a_k(X(t))\Delta t +o(\Delta t) =  \ &\hbox{the probability that
      reaction $k$}\\ 
    &\hbox{initiates in a small time interval $[t, t + \Delta t)$},
    \label{eq:FundPremdelay}
  \end{split}
\end{align}
where $o(\Delta t)/\Delta t \to 0$ as $\Delta t \to 0$.  Thus, no
matter whether a reaction is contained in $ND$, $CD$, or $ICD$, the
number of {\em initiations} at absolute time $t$ will be given by
\begin{equation}
  \hbox{number of initiations of reaction $k$ by time } t
  = Y_k\left( \int_0^t a_k(X(s)) ds\right),
  \label{eq:initiations}
\end{equation}
where the $Y_k$ are independent, unit rate Poisson processes.

Because the assumption \eqref{eq:FundPremdelay}, and hence equation
\eqref{eq:initiations}, only pertains to the initiation times of
reactions we must handle the completions separately.  There are three
different types of reactions, so there are three cases that need
consideration.

\vspace{.125in}

\noindent \textbf{Case 1}: If reaction $k$ is in $ND$ and initiates at
time $t$, then the system is updated by losing the reactant species
and gaining the product species at the time of initiation.

\vspace{.125in}

\noindent \textbf{Case 2}: If reaction $k$ is in $CD$ and initiates at
time $t$, then the system is updated only at the time of completion,
$t + \tau_k$, by losing the reactant species and gaining the product
species.

\vspace{.125in}

\noindent \textbf{Case 3}: If reaction $k$ is in $ICD$ and initiates
at time $t$, then the system is updated by losing the reactant species
at the time of initiation, $t$, and is updated by gaining the product
species at the time of completion, $t + \tau_k$.

\vspace{.125in}

\noindent The system can be written in the following integral form
\begin{align}
  \begin{split}
    X(t) = X(0) &+ \sum_{k \in ND} Y_{k}\left( \int_0^t a_k(X(s))ds
    \right)(\nu_k' - \nu_k)\\
    &+ \sum_{k \in CD} Z_{k}\left( \int_{0}^t a_k(X(s - \tau_j))ds
    \right)(\nu_k' - \nu_k)\\
    &+ \sum_{k \in ICD} W_{k}\left( \int_{0}^t a_k(X(s-\tau_k))ds
    \right)\nu_k' - \sum_{k \in ICD} W_{k}\left( \int_{0}^t
      a_k(X(s))ds \right)\nu_k,
  \end{split}
\end{align}
where each $a_k(s) = 0$ for $s < 0$, and the $Y_k$'s, $Z_k$'s, and
$W_k$'s are independent, unit rate Poisson processes.

We note that there are more potential cases than those listed above.
For example, the delay times, $\tau_k$, may best be described as a
random variable as opposed to being fixed or there could be multiple
completion times for a single initiation (implying things happen in
some order).  For the sake of clarity we do not consider such systems
in this paper but point out that it is a trivial exercise to extend
the results of this section to such systems.

\subsection{Current Algorithms}

Based upon the discussion above, we see that simulation methods for
systems with delays need to calculate when reactions initiate and
store when they complete.  However, because of the delayed reactions,
the propensity functions can change between initiation times.  Bratsun
et al. \cite{Bratsun2005} and Barrio et al. \cite{Barrio2006} used an
algorithm for computing the initiation times that is exactly like the
original Gillespie Algorithm except that if there is a stored delayed
reaction set to finish within a computed timestep, then the computed
timestep is discarded, and the system is updated to incorporate the
stored delayed reaction.  The algorithm then attempts another step
starting at its new state.  We will refer to this algorithm as the
Rejection Method.

\begin{algorithm}\textbf{(The Rejection Method)}
  \begin{enumerate}
    \setlength{\itemsep}{1pt}
    \setlength{\parskip}{0pt}
    \setlength{\parsep}{0pt}
  \item Initialize.  Set the initial number of molecules of each
    species and set $t = 0$.
  \item Calculate the propensity function, $a_k$, for each reaction.
    \label{RejReturn1}
  \item Set $a_0 = \sum_{k=1}^M a_k$.
  \item Generate an independent uniform(0,1) random number, $r_1$, and
    set $\Delta = 1/a_0\ln(1/r_1)$.
    \label{rejectRand}
  \item If there is a delayed reaction set to finish in $[t,t +
    \Delta)$
    \begin{enumerate}
      \setlength{\itemsep}{1pt}
      \setlength{\parskip}{0pt}
      \setlength{\parsep}{0pt}
    \item Discard $\Delta$.
      \label{discard}
    \item Update $t$ to be the time of the next delayed reaction,
      $\mu$.
    \item Update $x$ according to the stored reaction $\mu$.
    \item Return to step \ref{RejReturn1} or quit.
    \end{enumerate}
  \item Else
    \begin{enumerate}
      \setlength{\itemsep}{1pt}
      \setlength{\parskip}{0pt}
      \setlength{\parsep}{0pt}
    \item Generate an independent uniform(0,1) random number $r_2$.
    \item Find $\mu \in [1,\dots,m]$ such that
      \begin{equation*}
        \sum_{k=1}^{\mu-1} a_k < r_2a_0 \le \sum_{k=1}^{\mu} a_k,
      \end{equation*}
  \item If $\mu \in ND$, update the number of each molecular species
    according to reaction $\mu$.
  \item If $\mu \in CD$, store the information that at time $t +
    \tau_{\mu}$ the system must be updated according to reaction
    $\mu$.
  \item If $\mu \in ICD$, update the system according to the
    initiation of $\mu$ and store that at time $t + \tau_{\mu}$ the
    system must be updated according to the completion of reaction
    $\mu$.
  \item Set $t = t + \Delta$
  \item Return to step \ref{RejReturn1} or quit.
  \end{enumerate}
  \end{enumerate}
  \label{alg:rej}
\end{algorithm}


At first observation the statistics of the sample paths computed by
the above algorithm appear to be skewed because some of the timesteps
are discarded in step \ref{discard}. However, because the initiation
times are governed by Poisson processes via \eqref{eq:initiations}, we
may invoke the loss of memory property and conclude that the above
method is statistically exact.

The number of discarded $\Delta$'s will be approximately equal to the
number of delayed reactions that initiate.  This follows because,
other than the stored completions at the time the script terminates,
every delayed completion will cause one computed $\Delta$ to be
discarded.  Cai notes that the percentage of random numbers generated
in step \ref{rejectRand} and discarded in step \ref{discard} can
approach 50\%. \cite{Cai2007} Cai then develops an algorithm, called
the Direct Method for systems with delays, in which no random
variables are discarded.  We present Cai's Direct Method below,
however we refer the reader to Ref. \onlinecite{Cai2007} for full
details.

The principle of Cai's Direct Method is the same as that of the
original Gillespie Algorithm and the Rejection Method above: use one
random variable to calculate when the next reaction initiates and use
another random variable to calculate which reaction occurs at that
future time.  However, Cai updates the state of the system and
propensity functions due to stored delayed reactions during the search
for the next initiation time.  In this way he ensures that no random
variables are discarded as in the Rejection Method.

Suppose that at time $t$ there are ongoing delayed reactions set to
complete at times $t + T_1, t + T_2, \dots, t + T_d$.  Define $T_0 =
0$ and $T_{d+1} = \infty$.  According to Cai's Direct Method, in order
to calculate the time until the next reaction initiates, we first ask
if the reaction takes place before $t + T_1$.  If so, we may perform
the step.  If not, we must update the system according to the
completion of the reaction due to complete at time $t + T_1$, update
our propensity functions, and ask if the reaction takes place between
$t + T_1$ and $t + T_2$.  In this manner we will eventually find when
the next reaction initiates.  Following the lead of Cai, we first
present a method used for generating $\Delta$. \cite{Cai2007}

\begin{algorithm}\textbf{($\Delta$ generation for the Direct Method for
    systems with delays)}
  \begin{enumerate}
    \setlength{\itemsep}{1pt}
    \setlength{\parskip}{0pt}
    \setlength{\parsep}{0pt}
  \item Input the time $t$ and $a_0 = \sum_{k}a_k$.
  \item Generate an independent uniform(0,1) random number $r_1$.
  \item If no ongoing delayed reactions, set $\Delta =
    1/a_0\ln(1/r_1)$.
  \item Else
    \begin{enumerate}
      \setlength{\itemsep}{1pt}
      \setlength{\parskip}{0pt}
      \setlength{\parsep}{0pt}
    \item Set $i = 0$, $F = 0$, and $a_t = a_0T_1$.
    \item While $F < r_1$
      \begin{enumerate}
        \setlength{\itemsep}{1pt}
        \setlength{\parskip}{0pt}
        \setlength{\parsep}{0pt}
      \item Set $F = 1 - \hbox{exp}(-a_t)$.
      \item Set $i = i+ 1$.
      \item Calculate the propensity functions $a_k(t + T_i)$ due to the finish
        of the delayed reaction at $t + T_i$, and calculate $a_0(t+T_i)$.
      \item Set $a_t = a_t + a_0(t+ T_i)(T_{i+1} - T_i)$.
      \item If $i > 1$ update the state vector $x$ due to the finish
        of the delayed reaction at $t + T_{i-1}$.
      \end{enumerate}
    \item EndWhile
    \end{enumerate}
  \item Set $i = i - 1$.
  \item Set $\Delta = T_i - \left(\ln(1 - r_1) + a_t -
      a_0(t+T_i)(T_{i+1} - T_i) \right)/a_0(t+T_i)$.
  \item EndIf
  \end{enumerate}
  \label{alg:DeltaDirect}
\end{algorithm}

Because $T_1,\dots,T_d$ are needed to perform the simulation, Cai
introduces a $d \times 2$ matrix, $Tstruct$, whose $i$th row contains
$T_i$ and the index $\mu_i$ of the reaction due to complete at time $t
+ T_i$.  During a simulation, if we find that $\Delta \in
[T_i,T_{i+1})$, we delete rows 1 through $i$ of $Tstruct$ and set $T_j
= T_j - \Delta$ for all of the other delay times.  Also, rows are
added to $Tstruct$ when delayed reactions are initiated in such a way
that we always maintain $Tstruct(i,1) < Tstruct(i+1,1)$.  We present
Cai's direct method below.

\begin{algorithm}\textbf{(Direct Method for systems with delays)}
  \begin{enumerate}
    \setlength{\itemsep}{1pt}
    \setlength{\parskip}{0pt}
    \setlength{\parsep}{0pt}
  \item Initialize.  Set the initial number of molecules of each
    species and set $t = 0$. Clear $Tstruct$.
  \item Calculate the propensity function, $a_k$, for each reaction.
    \label{returnDirect}
  \item Set $a_0 = \sum_{k=1}^M a_k$.
  \item Generate $\Delta$ via Algorithm \ref{alg:DeltaDirect}.  If
    $\Delta \in [T_i,T_{i+1})$ update $Tstruct$ by deleting rows 1
    through $i$ and update the other delay times as described in the
    above paragraph.
    \label{Directstep}
  \item Generate an independent uniform(0,1) random number $r_2$.
  \item Find $\mu \in [1,\dots,m]$ such that
    \begin{equation*}
      \sum_{k=1}^{\mu-1} a_k < r_2 a_0 \le \sum_{k=1}^{\mu} a_k,
    \end{equation*}
    where the $a_k$'s and $a_0$ are generated in step
    \ref{Directstep}.
  \item If $\mu \in ND$, update the number of each molecular species
    according to reaction $\mu$.
  \item If $\mu \in CD$, update $Tstruct$ by adding the row
    $[\tau_{\mu} \ , \ \mu]$ so that $Tstruct(i,1) < Tstruct(i+1,1)$
    still holds for all $i$.
  \item If $\mu \in ICD$, update the system according to the
    initiation of $\mu$ and update $Tstruct$ by adding the row
    $[\tau_{\mu} \ , \ \mu]$ so that $Tstruct(i,1) < Tstruct(i+1,1)$
    still holds for all $i$.
  \item Set $t = t + \Delta$.
  \item Return to step \ref{returnDirect} or quit.
  \end{enumerate}
  \label{alg:cai}
\end{algorithm}

We note that the Direct Method will use precisely one random number to
find each initiation time.  In this way the Direct Method is more
efficient than the Rejection Method, which discards a $\Delta$ (and
therefore a random number) each time a delayed reaction completes.
However, the extra machinery built into the Direct Method in order to
find $\Delta$ will slow the algorithm as compared with the Rejection
Method.  Therefore, it is not immediately clear which method will
actually be faster on a given system.

\subsection{The modified Next Reaction Method for systems with delays}

We now extend our modified Next Reaction Method to systems with
delays.  Recall that the central idea behind the modified Next
Reaction Method is that knowledge of the internal time at which $Y_k$
fires next can be used to generate the absolute time of the next
initiation of reaction $k$.  The same idea works in the case of
systems with delays because the initiations are still given by the
firing times of independent Poisson processes via equation
\eqref{eq:initiations}.  Therefore, if $T_k$ is the current internal
time of $Y_k$, $P_k$ the first internal time after $T_k$ at which
$Y_k$ fires, and the propensity function for the $k$th reaction
channel is given by $a_k$, then the time until the next initiation of
reaction $k$ (assuming no other reactions initiate or complete) is
still given by $\Delta t_k = (P_k - T_k)/a_k$.  The only change to the
algorithm will be in keeping track and storing the delayed
completions.  To each delayed reaction channel we therefore assign a
vector, $s_k$, that stores the completion times of that reaction in
ascending order.  Thus, the time until there is a change in the state
of the system, be it an initiation or a completion, will be given by
\begin{equation*}
  \Delta = \min\{\Delta t_k,s_k(1)-t\},
\end{equation*}
where $t$ is the current time of the system.  These ideas form the
heart of our Next Reaction Method for systems with delays:

\begin{algorithm} \textbf{(Next Reaction Method for systems with
    delays)}
  \begin{enumerate}
    \setlength{\itemsep}{1pt}
    \setlength{\parskip}{0pt}
    \setlength{\parsep}{0pt}
  \item Initialize.  Set the initial number of molecules of each
    species and set $t = 0$.  For each $k \le M$, set $P_k = 0$ and
    $T_k = 0$, and for each delayed reaction channel set $s_k = [\infty]$.
  \item Calculate the propensity function, $a_k$, for each reaction.
  \item Generate $M$ independent, uniform(0,1) random numbers, $r_k$,
    and set $P_k = \ln(1/r_k)$.
  \item Set $\Delta t_k = (P_k - T_k)/a_k$.
    \label{impstepDelay}
  \item Set $\Delta = \min_k\{\Delta t_k,s_k(1)-t\}$.
  \item Set $t = t + \Delta$.
  \item If we chose the completion of the delayed reaction $\mu$:
    \begin{itemize}
      \setlength{\itemsep}{1pt}
      \setlength{\parskip}{0pt}
      \setlength{\parsep}{0pt}
    \item Update the system based upon the completion of the reaction
      $\mu$.
    \item Delete the first row of $s_{\mu}$.
    \end{itemize}
  \item Elseif reaction $\mu$ initiated and $\mu \in ND$
    \begin{itemize}
      \setlength{\itemsep}{1pt}
      \setlength{\parskip}{0pt}
      \setlength{\parsep}{0pt}
    \item Update the system according to reaction $\mu$.
    \end{itemize}
  \item Elseif reaction $\mu$ initiated and $\mu \in CD$
    \begin{itemize}
      \setlength{\itemsep}{1pt}
      \setlength{\parskip}{0pt}
      \setlength{\parsep}{0pt}
    \item Update $s_{\mu}$ by inserting $t + \tau_{\mu}$ into $s_{\mu}$ in
      the second to last position.
    \end{itemize}
  \item Elseif  reaction $\mu$ initiated and $\mu \in ICD$
    \begin{itemize}
      \setlength{\itemsep}{1pt}
      \setlength{\parskip}{0pt}
      \setlength{\parsep}{0pt}
    \item Update the system based upon the initiation of reaction
      $\mu$.
    \item Update $s_{\mu}$ by inserting $t + \tau_{\mu}$ into $s_{\mu}$ in
      the second to last position.
    \end{itemize}
  \item For each $k$, set $T_k = T_k + a_k\Delta$.
  \item If reaction $\mu$ initiated, let $r$ be uniform(0,1) and set
    $P_{\mu} = P_{\mu} + \ln(1/r)$.
  \item Recalculate the propensity functions, $a_k$.
  \item Return to step \ref{impstepDelay} or quit.
  \end{enumerate}
  \label{alg:nrmdelay}
\end{algorithm}

We note that after the first step, the Next Reaction Method for
systems with delays only generates one random variable for each
initiation as opposed to the two generated in the Direct Method.
Further, Algorithm \ref{alg:nrmdelay} performs the updates in a way
that uses every random variable that is calculated yet does not have
the complicated machinery necessary in the Direct Method.  We should
therefore expect that Algorithm \ref{alg:nrmdelay} will need less time
in the simulation of chemical reaction systems with delays then either
the Rejection or Direct Method.  We also note that similar to our
modified Next Reaction Method, Algorithm \ref{alg:nrmdelay} extends
easily to systems with time dependent rate constants, and
non-exponential waiting times between initiations.

\subsection{Numerical examples}

\begin{example}

Consider the following
system consisting of two reaction channels:
\begin{equation}
  R_1: \  X_1 + X_2 \overset{c_1}\to X_3 \hspace{.5in} R_2: \  X_3
  \overset{c_2}\to \emptyset .
  \label{delaysystem}
\end{equation}
The reaction channel $R_1$ belongs to $ICD$ and $R_2$ belongs to $ND$.
Therefore, we update $X_1 = X_1 - 1$ and $X_2 = X_2 - 1$ at the moment
of initiation of $R_1$, but only update $X_3 = X_3 + 1$ after a delay.
Following Cai, \cite{Cai2007} we chose $c_1 = 0.001$, $c_2 = 0.001$,
$X_1(0) = 1000,$ $X_2(0) = 1000$ and $X_3(0) = 0$. We let the delay of
$R_1$ be $\tau_1 = 0.1$ and simulated this system from time $t = 0$
until $t = 1$.  These values were chosen so that the number of
initiations that have delayed completions is approximately 100\% of
all initiations.  Therefore, nearly 50\% of all steps of the Rejection
Method will discard a random variable, thereby maximizing its
wastefulness.

We performed $10^4$ simulations using each of the Rejection, Direct,
and Next Reaction Method for systems with delays.  The Rejection
Method of Barrio and Bratsun took 179.5 CPU seconds, the Direct Method
of Cai took 167.2 CPU seconds, and the Next Reaction Method took 82.8
CPU seconds.  Therefore, the Rejection Method took 7.4\% more time
than the Direct Method and took 116.8\% more time than our Next
Reaction Method for systems with delays while the Direct Method took
101.9\% more time than our Next Reaction Method.  We note that we have
not reproduced the results stated in Ref. \onlinecite{Cai2007} where
the Direct Method was found to be 23\% more efficient than the
Rejection Method.  In fact, when the Direct and Rejection Methods are
programed in such a way that the differences in the codes reflects the
differences in the algorithms, one typically finds that the difference
in simulation times does not differ substantially.  Considering that
for this example nearly half of all random numbers generated by the
Rejection method in order to calculate $\Delta$ are discarded (which
is a maximum in waste for the Rejection Method, see Ref.
\onlinecite{Cai2007}), the fact that the Direct Method is not
substantially more efficient than the Rejection Method points out that
the time used by the steps in the Direct Method in order to calculate
$\Delta$ is not negligible as compared to the time needed to generate
random numbers.
\end{example}

Because the Rejection Method becomes more wasteful as the number of
rejected $\Delta$'s increases, we will test the three algorithms on a
system in which we can easily control the percentage of $\Delta$'s
that are discarded.

\begin{example}

  We consider a simple model of gene transcription whose non-delayed
  version can be found in Ref. \onlinecite{Rempala06}.  The model
  consists of three species: gDNA (NN), messenger RNA (mRNA), and the
  catalytic TProt.  NN is assumed to be in such abundant quantities as
  to be constant, so the model is completely determined by the state
  of the species mRNA and TProt.  There are four reactions allowed in
  the model:
  \begin{equation}
    \begin{array}{ll}
      R_1: \ \ NN  \xrightarrow{k_1TProt} mRNA \hspace{.5in}& R_3:\ \ 0
      \xrightarrow{k_3} TProt\\
      R_2: \ \ mRNA \xrightarrow{k_2} 0 & R_4:\ \ TProt \xrightarrow{k_4}
      0.
    \end{array}
    \label{examplesystem}
  \end{equation}
  We suppose that reaction one belongs to $CD$ and has a delay of
  $\tau = 5$. It is simple to show that the mean value of the state of
  the system has an equilibrium value of $(\overline {mRNA}, \overline
  {TProt}) = ((k_1k_3)/(k_2k_4), k_3/k_4)$, and the mean values of the
  propensities of the reactions have equilibrium values of
  \begin{equation*}
    \begin{array}{ll}
      \bar \lambda_1 = k_1\overline{TProt} = k_1 \frac{k_3}{k_4}
      \hspace{.5in} &   \bar \lambda_3 = k_3 \\
      \bar \lambda_2 = k_2\overline{mRNA} = k_1 \frac{k_3}{k_4} & \bar
      \lambda_4 = k_4 \overline{TProt} = k_3. 
    \end{array}
  \end{equation*}
  Therefore, the expected percentage of the initiations that have
  delayed completions can be approximated by $\gamma$, which is given
  by
  \begin{equation}
    \gamma = \frac{\bar \lambda_1}{\bar \lambda_1 + \bar \lambda_2 +
      \bar \lambda_3 +\bar \lambda_4} = \frac{k_1 \frac{k_3}{k_4}}{2k_1
      \frac{k_3}{k_4}+2k_3}=
    \frac{1}{2}\frac{\frac{k_1}{k_4}}{\frac{k_1}{k_4} + 1}.
    \label{eq:gamma}
  \end{equation}
  For the Rejection Method, the number of discarded $\Delta$'s will be
  approximately the number of initiations of delayed reactions.
  Therefore the Rejection Method becomes more wasteful as the
  percentage of the total reaction initiations that have delayed
  completions increases, and so we may expect to see that as $\gamma$
  increases the Direct Method will become relatively faster as
  compared to the Rejection method. To test this we set $k_2 = 1$,
  $k_3 = 15$, and $k_4 = 1$ so that $\gamma = (1/2)k_1/(k_1 + 1).$
  $k_1$ now acts as a parameter that can be changed in order to see
  the effect $\gamma$ has on the relative speeds of the two
  algorithms.  We note that the parameters were not chosen for their
  biological relevance, but instead were chosen for experimental ease.

  For a series of $k_1$'s we computed the CPU time needed for the
  Direct Method, Rejection Method, and Next Reaction Method for
  systems with delays to simulate the above system $10^4$ times from
  time 0 to time 30. See Figure \ref{fig:plot}.
  \begin{figure}
     \begin{center}
       \includegraphics[width = 5in]{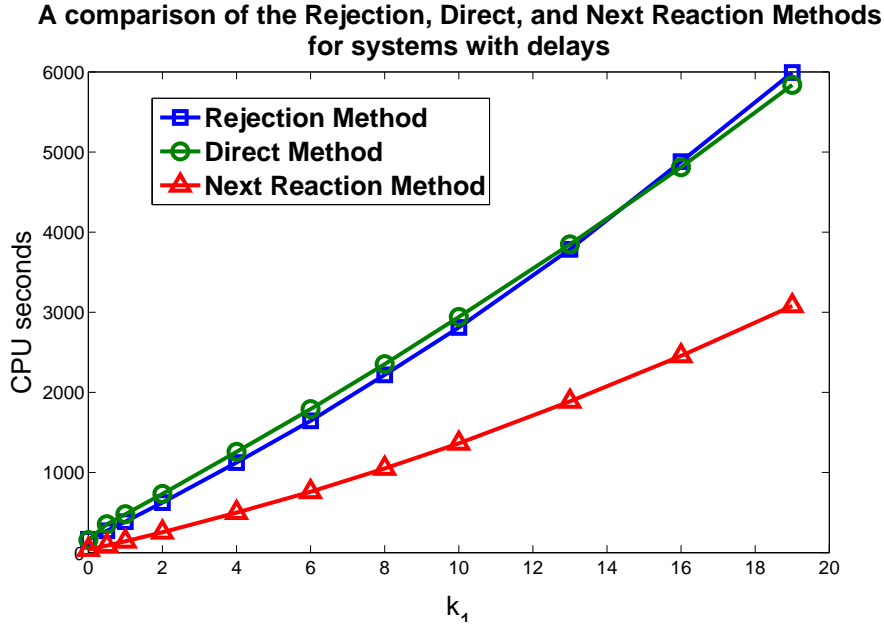}
     \end{center}
     \caption{The above plot compares the speeds of the Rejection
       Method, Direct Method, and Next Reaction Method for systems
       with delays as the percentage of timesteps that are rejected
       in the Rejection Method, as parameterized by $k_1$, increases.
       For different values of $k_1$, each method was used to
       simulate the system \eqref{examplesystem} $10^4$ times.  The
       plot above gives the CPU time needed for each method as a
       function of $k_1$. We see that the Rejection and Direct
       Methods are nearly equivalent while the Next Reaction Method
       for systems with delays is significantly more efficient than
       both for all $k_1$.}
    \label{fig:plot}
  \end{figure}
  We see that as $k_1$ increases, the Rejection and Direct Methods
  remain relatively close in terms of efficiency with the Rejection
  Method being slightly more efficient for smaller $k_1$ and slightly
  less efficient for larger $k_1$.  However, the Next Reaction Method
  for systems with delays (Algorithm \ref{alg:nrmdelay}), is
  significantly more efficient than both for all $k_1$.

\end{example}

\section{Conclusion}

By explicitly representing the reaction times of discrete stochastic
chemical systems with the firing times of independent, unit rate
Poisson processes with internal times given by integrated propensity
functions we have developed a modified Next Reaction Method.  We
extended our modified Next Reaction Method to systems with delays and
demonstrated its computational efficiency on such systems over the
Rejection Method of Bratsun et al. and Barrio et al., and the Direct
Method of Cai.  Considering that many models of natural cellular
processes such as gene transcription and translation have delays
between the initiation and completion of reactions, and that the
Rejection method appears to be the most widely used method for
simulating such systems, we feel that this extension will be useful.
Also, as is pointed out in the text, our modified Next Reaction Method
can be easily extended to systems with non-exponential waiting times
between initiations and is preferable to both the Gillespie Algorithm
and the original Next Reaction Method for systems with propensities
that depend explicitly on time.  We feel that having a single,
efficient simulation method applicable to such a broad range of
chemical systems will prove to be a beneficial contribution.

\begin{acknowledgments}
I would like to thank Thomas G. Kurtz for introducing me to the notion
of representing the reaction times of chemical systems with the firing
times of independent, unit rate Poisson processes undergoing random
time changes and for making the connection between this work and the
theory of generalized semi-Markov processes.  I would also like to
thank an anonymous reviewer for making several suggestions that
improved the clarity of this work.  This work was done under the
support of NSF grant DMS-0553687.
\end{acknowledgments}

\appendix

\section{Unfinished calculation}
\label{app:A}

In Section \ref{sec:time-dep} we showed that if a system has
propensity functions that depend explicitly on time, then the amount
of absolute time, $\Delta$, that must pass after time $t$ before any
reaction fires has distribution function
\begin{equation*}
  1 - \exp \left( - \sum_{k = 1}^M \int_t^{t + \Delta} a_k(X(t),s) ds \right).
\end{equation*}
where $r$ is uniform$(0,1)$. We will sketch the proof of why the
reaction that fires at that time will be chosen according to the
probabilities $a_k(X(t),t+\Delta)/a_0$, where $a_0 = \sum_{k=1}^M
a_k(X(t),t+\Delta)$.  

Let $H(r) \ \dot = \ \sum_{k = 1}^M \int_t^{t + r} a_k(X(t),s) ds.$
For $j \le M$, let $\Delta t_j$ be the amount of time that must pass
after time $t$ before the $j$th reaction fires.  Let $F$ denote the
random variable $\min\{\Delta t_j\}$.  Then, conditioning on the fact
that $F = \Delta$ and using the independence of the underlying Poisson
processes we have
\begin{align}
  \begin{split}
    P(\Delta t_k < \Delta t_{j,j \ne k} | F = \Delta) &= \lim_{\delta
      \to 0} P(\Delta t_k < \Delta t_{j,j\ne k} | F \in [\Delta,
    \Delta + \delta))\\
    &= \lim_{\delta \to 0}\frac{P(\Delta t_k < \Delta t_{j,j\ne k}, F
      \in [\Delta, \Delta +
      \delta))}{P(F \in [\Delta, \Delta + \delta))}\\
    &= \lim_{\delta \to 0}\frac{P(\Delta t_k \in [\Delta, \Delta +
      \delta), \Delta t_{j,j\ne k} > \Delta +
      \delta)}{\exp(-H(\Delta))
      - \exp(-H(\Delta + \delta))}\\
    &= \lim_{\delta \to 0}\frac{P(\Delta t_k \in [\Delta, \Delta +
      \delta))\prod_{j \ne k} P(\Delta t_{j} > \Delta +
      \delta)}{\exp(-H(\Delta)) - \exp(-H(\Delta + \delta))}.
  \end{split}
  \label{eq:long}
\end{align}
It is a simple exercise to show that for any $j \le M$
\begin{equation}
  P(\Delta t_j > s) = \exp\left(-\int_t^{t+s} a_j(X(t),s)ds\right).
  \label{eq:short}
\end{equation}
Combining equations \eqref{eq:long} and \eqref{eq:short} with an
application of L'Hopital's rule gives the desired result.

\bibliography{revtexAnd}

\pagebreak

\noindent \textbf{Figures}%

\vspace{.2in}

\noindent Figure 1. 
\vspace{1in}

    \begin{center}
       \includegraphics[width = 6in]{comp.eps}
     \end{center}

\pagebreak

\noindent \textbf{Captions}

\vspace{.2in}

\noindent Caption for Figure 1. 

\noindent The above plot compares the speeds of the Rejection Method,
Direct Method, and Next Reaction Method for systems with delays as the
percentage of timesteps that are rejected in the Rejection Method, as
parameterized by $k_1$, increases.  For different values of $k_1$,
each method was used to simulate the system \eqref{examplesystem}
$10^4$ times.  The plot above gives the CPU time needed for each
method as a function of $k_1$. We see that the Rejection and Direct
Methods are nearly equivalent while the Next Reaction Method for
systems with delays is significantly more efficient than both for all
$k_1$.

\end{document}